\documentclass[a4paper]{jpconf}
\usepackage{graphicx}
\usepackage{comment}
\usepackage{amsmath,amssymb,amsfonts,graphicx,dcolumn}
\usepackage{floatrow}
\usepackage{epsfig}
\usepackage{wrapfig}
\usepackage{lipsum}
\usepackage{setspace}

\begin{document}
\title{Nucleon 3D structure from double parton
scattering: a Light-Front quark model
analysis}

\author{Matteo Rinaldi}

\address{Instituto de Fisica Corpuscular (CSIC-Universitat de Valencia), 
Parc Cientific UV, C/ Catedratico Jose Beltran 2, E-46980 Paterna 
(Valencia), Spain.
}

\ead{mrinaldi@ific.uv.es;  matteo.rinaldi@pg.infn.it}

\begin{abstract}
Double parton distribution functions (dPDFs) represent a tool to explore
the 3D partonic  structure of the proton. They can be measured in
high energy
proton-proton and proton nucleus collisions and
encode information on how partons inside a proton are correlated among each
other. dPFDs are studied here
in the valence region by
means of a constituent quark model scenario within the relativistic Light Front 
approach, where two particle
correlations are present without any additional
prescription. 
 Furthermore, a  study of the QCD evolution at high energy scale, of the
model results, has been  completed in order to compare our predictions  with 
future data analyses. In closing, 
results on the evaluation  of the so called $\sigma_{eff}$, crucial ingredient 
for the description of double parton scattering, where dPDFs can be accessed, 
are presented and discussed.  

\end{abstract}

\section{Introduction}
The observation of multiple parton interactions (MPI), occurring
in high energy hadron-hadron collisions, will be an important goal of the 
 experimental and theoretical analyses of the processes studied, e.g., at 
the LHC. 
In particular, in these kind of events, more than one parton of a hadron 
interact with  partons of the other colliding hadron. Naturally,
 MPI contribution  to the total cross section is  
suppressed  with respect to the single parton interaction.
However, the measurement and estimate of MPI cross sections  is an important
 challenge being MPI events  a background for the search of new  Physics.
 Furthermore, from a hadronic point of view,  we are interested on MPI due to 
the 
possibility of accessing new fundamental information on the partonic proton 
structure. We focus our studies on the  the double parton 
scattering (DPS), the most simple case of MPI, which can be 
observed, in principle, 
 in several processes, e.g., $WW$ with dilepton productions
and double
Drell-Yan processes (see, Refs. \cite{3a,4a,5a,6a, report} for
recent 
reviews).
At the LHC, DPS, has been
observed some years ago \cite{16a} and represents  a
background for the Higgs production in several channels.
From a theoretical point of view, 
the DPS cross section
is written in terms of a new quantity,  the so called double 
parton distribution function (dPDF),
$F_{ij}(x_1,x_2,{\vec z}_\perp,\mu)$, which describe the 
joint probability 
of finding two partons of flavors $i,j=q, \bar q,g$ with 
longitudinal momentum fractions $x_1,x_2$ and separation 
$\vec z_\perp$ in the transverse plane inside the hadron, see Ref. \cite{1a}.
Here $\mu$ is the renormalization scale.
However, to date, being  dPDFs  poorly known, in order to 
qualitatively estimate the magnitudo  of the DPS cross section, 
the following approximation
is usually adopted: 

\vskip -0.4cm
\begin{eqnarray} 
\label{app}
 \hskip -3mm F_{ij}(x_1,x_2,\vec z_\perp,\mu) &=& q_{i}(x_1,\mu)~
q_{j}(x_2,\mu)~T(\vec z_\perp,\mu) 
 \theta(1-x_1-x_2) (1-x_1-x_2)^n
~,
\end{eqnarray}
\vskip -0.1cm
{ i.e.}, dPDFs are
evaluated in a fully factorized ansatz in terms of standard 
 one-body parton distribution functions (PDF), $q(x)$, and  $T(\vec 
z_\perp,\mu)$, the function encoding 
 all parton 
correlations in the transverse plane.
The spirit of the latter assumption is to neglect all possible 
 double parton
correlations (DPCs) between the two interacting partons, being the latter 
almost unknown, see e.g.  Ref. \cite{kase1} for updates.
Moreover
 dPDFs are non
perturbative quantities in QCD so that they cannot be easily evaluated from
the theory.
Nevertheless, as already deeply discussed in Refs. \cite{man,noi1,noij1,bru}, 
quark model calculations of dPDFs could help to grasp the basic feature of such 
quantities.
This strategy has been largely used in the past to study unknown distributions. 
However, since in this scenario 
 dPDFs are calculated at the low
hadronic scale of the model, $\mu_0 \sim \Lambda_{QCD}$, in order
 to compare the obtained outcomes with future data taken at
high energy scales, $Q > \mu_0$, it is then necessary to perform the
perturbative QCD (pQCD) evolution of the model calculations, 
using the dPDF evolution equations, see Refs. \cite{23a,24a}. 
The idea supporting our analysis is that,
thanks to this procedure, future data analyses
of the DPS processes could be guided, in principle, by 
model calculations.
In the first part of the present paper, we focus our attention on the study  of 
the
role 
 of DPCs 
 in dPDFs in
order to verify, as a first step, 
the validity of the approximation Eq. (\ref{app}),
often used for  data analyses. In particular, it is here recall 
that 
in all model calculations of dPDFs, see Refs. \cite{man,noi1,noij1},
 the
assumption  Eq. (\ref{app}) is  violated. See also Ref. 
\cite{noice} for details on the violation of the factorized ansatz due to model 
independent 
relativistic effects in the calculated dPDFs.
Moreover, as already pointed out, it is fundamental to realize to which 
extend DPCs survive at very high energy scales, where, due to the large 
population of partons, the role of DPCs  could be less important than 
at the low 
scale of the used model.
To this aim in Refs. \cite{noij1, noij2}, DPCs in dPDFs have been studied at  
the energy 
scale of the experiments. In particular in Ref. \cite{noij2} 
 an extension of the approach  used  in Refs. \cite{noij1} has been 
provided to include sea quarks and gluon degrees of freedom. 
Here, the most important results of the these analyses will be summarized.
In the last section of the present paper, an investigation of the so called 
$\sigma_{eff}$ will be  discussed. Usually, indeed, DPS cross section, in 
processes with final 
state 
$A+B$, is written through   the following ratio  (see 
e.g.~Ref.~\cite{MPI15}):
\vskip -0.4cm
\begin{eqnarray}
\sigma^{A+B}_{DPS}  = \dfrac{m}{2} \dfrac{\sigma_{SPS}^A 
\sigma_{SPS}^B}{\sigma_{eff}}\,,
\label{sigma_eff_exp}
\end{eqnarray}
\vskip -0.2cm
where $m$ is combinatorial factor depending on the final states $A$ and $B$ 
($m=1$ for $A=B$ or $m=2$ for $A \neq B$) and
$\sigma^{A(B)}_{SPS}$ is the single parton scattering cross section with final 
state $A(B)$. {Expressing the $\sigma_{DPS}$ cross section in 
Eq.~(\ref{sigma_eff_exp})
in terms of product of $\sigma_{SPS}$, one assumes that, as a first 
approximation,  double parton 
distributions can be written as in Eq. (\ref{app}).
The present knowledge on DPS cross sections
has been condensed in the experimental and model dependent extraction of 
$\sigma_{eff}$~\cite{MPI15,S1,S2,S3,S4,S5,S6,S7}. 
To date, $\sigma_{eff} \simeq$ 
15 mb, compatible, within errors, 
with a constant, irrespective of centre-of-mass energy of the hadronic 
collisions, the final state  and $x_i$. Let us stress again that the latter 
result is obtained considering the fully uncorrelated scenario shown in Eq. 
(\ref{app}) where all DPCs are neglected.
Also in this case non perturbative methods have been used to study
$\sigma_{eff}$, through the calculation of dPDFs, in order to characterize 
``signals'' of DPCs in $\sigma_{eff}$. On top of that, in Ref. \cite{noiPLB}, 
the   possible dependence of the latter 
quantity on the longitudinal momentum fraction carried by the interacting 
partons has been addressed. Recent results on this topic will be here 
discussed, including also new intriguing outcomes addressed in Refs. 
\cite{noice, noiads}. In Ref. \cite{noice} the contribution of relativistic 
effects to 
$\sigma_{eff}$ have been estimated by means of the  Light-Front 
approach (LF), already used for the evaluation of dPDFs in Ref. \cite{noij1}. 
Furthermore, in Ref. \cite{noiads}, $\sigma_{eff}$ has been calculated in a 
AdS/QCD framework, showing, also here,  a strong $x_i$ dependence of 
$\sigma_{eff}$ 
in the valence region.

\section{Calculation of dPDFs  within LF CQMs}
In this section, details of the calculations of dPDFs,
within the LF approach (see Refs. \cite{pol,pauli} for 
fundamental details),
will be discussed verifying  if  the factorized
ansatz Eq. (\ref{app})
would work in this relativistic scenario.
In this framework, among all positive features of the LF approach, let us 
remark that
 one can
obtain a fully Poincar\'e covariant description of relativistic
strongly interacting systems with a fixed number of on-shell constituents,
 LF boosts and the ``plus'' components of momenta ($a^+ = a_0 +
a_3$) are kinematical operators. Furthermore, being the LF hypersurface, the 
one where
the initial conditions of the system are fixed, tangent to the light-cone,
the kinematics of DIS processes is naturally obtained. For these reasons, this 
strategy has been adopted in the past for the evaluation unknown distributions, 
see 
e.g. Refs. \cite{boffi1,boffi3,traini14}.
All the details on the present calculations are deeply discussed and reported 
in Ref. \cite{noij1} and will  not be repeated here.
The final expression of the dPDF in momentum space for two unpolarized quarks 
of flavor 
$u$,  reads:
\vskip -0.5cm
\begin{eqnarray}
\label{main}
\nonumber
\hskip -9mm  F_{uu}^{\lambda_1,\lambda_2}(x_1, x_2, \vec k_\perp) 
& = & 
2(\sqrt{3})^3 \int
\left[
\underset{i=1}{\overset{3}\prod} d \vec k_i 
\underset{\lambda_i^f \tau_i} {\sum}
\right]
\delta \left(
\underset{i=1}{\overset{3}\sum} \vec k_i 
\right) 
\Psi^* \left(\vec k_1 +
\dfrac{\vec k_\perp}{2}, \vec k_2 -
\dfrac{\vec k_\perp}{2},\vec k_3
; \{\lambda_i^f, \tau_i \}
\right) \\
\nonumber
& \times & 
\Psi \left(\vec k_1 -
\dfrac{\vec k_\perp}{2}, \vec k_2 +
\dfrac{\vec k_\perp}{2}, \vec k_3 
; \{\lambda_i^f, \tau_i \}
\right)
 \delta \left(x_1 
-\dfrac{k_1^+}{P^+}
\right) \delta \left(x_2 -\dfrac{k_2^+}{P^+}
\right)~,
\end{eqnarray}
\vskip -0.3cm
being 
$k_\perp$  the relative transverse momentum of one of the parton in the 
amplitude and its complex conjugate and $\vec k_i$ the momentum of the $i-$ 
quark.

The canonical proton wave function $\psi^{[c]}$ is embedded in
the function
$\Psi$ here above, which can be written as follows:

\vskip -0.5cm
\begin{eqnarray}
\label{psiint}
\hskip -3mm \Psi(\vec k_1,\vec k_2,\vec k_2; \lbrace \lambda_i^f, \tau_i
\rbrace) &=&
\underset{i=1}{\overset{3} \prod} \left [
\underset{\lambda_i^c}\sum D^{*1/2}_{\lambda_i^c
\lambda_i^f}(R_{cf}(\vec k_i))
\right ]
 \psi^{[c]}(\{\vec k_i, \lambda_i^c, \tau_i   \} )~,
\end{eqnarray}
\vskip -0.2cm
where $\lambda_i^c$ and $\tau_i$ are the canonical parton helicity and
the isospin, respectively.
Here the short notation $\{\alpha_i\}=\alpha_1,\alpha_2,\alpha_3 $ is 
introduced.
In Eq. (\ref{psiint}), the Melosh operators, which allow to rotate the LF spin 
into the canonical one are introduced:
\vskip -0.4cm
\begin{eqnarray}
\label{mel1}
\hskip -9mm  \hat D_i = D^{1/2}_{\mu
\lambda}(R_{cf}(\vec k_i)) =
\langle \mu  \left|~ \dfrac{m +x_iM_0 - i \vec 
\sigma_i
\cdot (\hat z_i \times \vec k_{i\perp})}{\sqrt{(m +x_i M_0)^2+\vec
k_{i\perp}^2}}
~\right| \lambda \rangle~,
\end{eqnarray}
\vskip -0.4cm
being $M_0 = \underset{i}{\sum} \sqrt{m^2+ \vec k_i^2} $  the total
free energy mass of the  system and $\mu$ and $\lambda$ 
generic canonical spins.
In this scenario, being $M_0$, appearing in delta functions in Eq. 
(\ref{main}), 
 a kinematical quantity, dPDFs are well defined in the $x_1+x_2<1$ region,
at variance with
what happens in the instant form 
calculations of PDFs and  dPDFs (see, e.g., Ref. \cite{noi1}). 
Let us remark  that the main ingredient in Eq. (\ref{main})
is the canonical proton wave function which has been calculated by means of
 a
relativistic CQM, the so called hyper-central CQM described in Ref. \cite{49a}. 
The choice
of this model is motivated by its simplicity and capability to basically
reproduce  the spectrum of light-hadrons. 
 Since actually for dPDFs  there are not yet available data, model calculations 
of these quantities could provide
essential information on their relevant features.
Before discussing in the next section the results of the  calculations of 
dPDFs, one should notice that  in principle
 the role of the model independent Melosh 
rotations  can be important. Indeed, from Eq. (\ref{mel1}), one can 
realize that such 
operator con introduce, in the calculation of dPDFs, non trivial correlations 
between $x_i$ 
and $k_\perp$. As deeply discussed in Ref. \cite{noice}, in order to taste 
possible effects of these objects, it is instructive to study 
the following quantity:
\vskip -0.4cm
\begin{eqnarray}
\label{Mel}
DD^\dagger (\vec k_\perp,x_1,x_2, \vec k_{1\perp}=0, \vec k_{2\perp}=0  ) = 
\langle SU(6) | D_1^\dagger D_1  D_2^\dagger D_2 |SU(6)\rangle ~,
\end{eqnarray}
\vskip -0.2cm
being $\vec 
k_{i \perp}$ the intrinsic transverse component of the  $i$ 
quark momentum. 
In the latter quantity use has been made of the commonly adopted SU(6) symmetry 
in order to 
evaluate the spin part of the proton wave function. In this scenario, the 
quantity (\ref{Mel}) is rather model independent.
Here we consider 
two fast partons (FF) with $x_1=0.2,~x_2=0.3$,
one slow and and one fast parton (SF) with $x_1=0.04,~x_2=0.3$ and
two slow partons (SS) with $x_1=0.04,~x_2=0.03$.
The calculation of Eq.~(\ref{Mel}) in these kinematic regions is
presented in Fig.~(\ref{fig1})
where one may identify three distinct regions as a function of $k_\perp$.
For $k_\perp \rightarrow 0$ the Melosh's in all kinematic configurations 
reduce to unity. In an intermediate region 
of $k_\perp$ the curves show a dip whose depth depends on the chosen kinematic 
configuration and 
at larger $k_\perp$ the curves flattens out with different asymptotics. 
This complicated pattern, generated by Melosh's rotations, affects
the calculation of dPDFs, which, in general, are distributions
evaluated also at $k_\perp \neq 0$.

\begin{wrapfigure}{L}{0.5\textwidth}
\begin{center}
\centering
{\includegraphics[width=16pc]{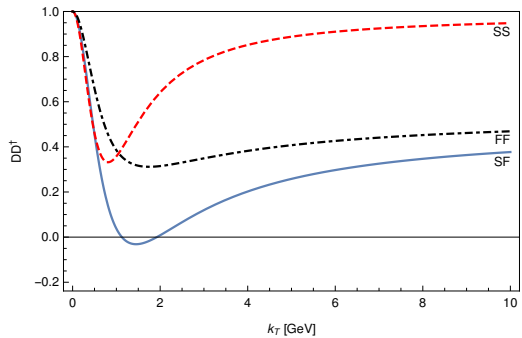}}
\end{center}
 {\caption{\footnotesize{ The quantity  Eq.~(\ref{Mel}), as 
function 
of 
$k_T= 
k_\perp $,  evaluated in different regions of $x_1$ and $x_2$ 
with $\vec k_{1\perp}=\vec k_{2\perp}=0$}.}\label{fig1}}
\end{wrapfigure}
\noindent

\section{Results of the calculations of dPDFs at the hadronic scale}

In this section the main results of the calculations of  dPDFs at the 
hadronic scale,
$\mu_0^2 \sim 0.1$ GeV$^2$, will
be presented. In particular, as already mentioned, here the emphasis of the 
 analysis is 
focused on
testing the validity of the approximation Eq. (\ref{app}).
As already discussed in the previous section, in general, Melosh rotations 
introduce correlations between $k_\perp$
and $x_1, x_2$ so that, as also found in Ref. \cite{noij1}, at the hadronic 
scale such factorization is violated in all model calculations of dPDFs within 
the LF approach, see Ref. \cite{noice}. 
Furthermore, 
in order  to study the validity of the approximations Eq.
(\ref{app}),  in the $x_1$ and $x_2$   dependence of dPDFs, the following 
ratio have been evaluated:
\vskip -0.4cm
\begin{eqnarray}
\label{ratio12u}
 r_2 = 
\dfrac{2 uu(x_1,x_2, k_\perp=0)}{u(x_1)u(x_2)}~,
\end{eqnarray}
\vskip -0.2cm
 As one can 
see on the left panel of Fig. (\ref{5}), 
being the ratio $r_2$, Eq. (\ref{ratio12u}), different from the unity in all 
the kinematical range,
also  the
factorized form of dPDFs in terms of the product of single PDFs is not 
supported by 
the present
approach.  In closing, at the hadronic scale, in all model calculations of 
dPDFs (see also Refs. \cite{man, noi1}), the assumption Eq. (\ref{app}) is 
violated due to DPCs.

\begin{figure}[b]
\vspace{6.8cm}
\includegraphics{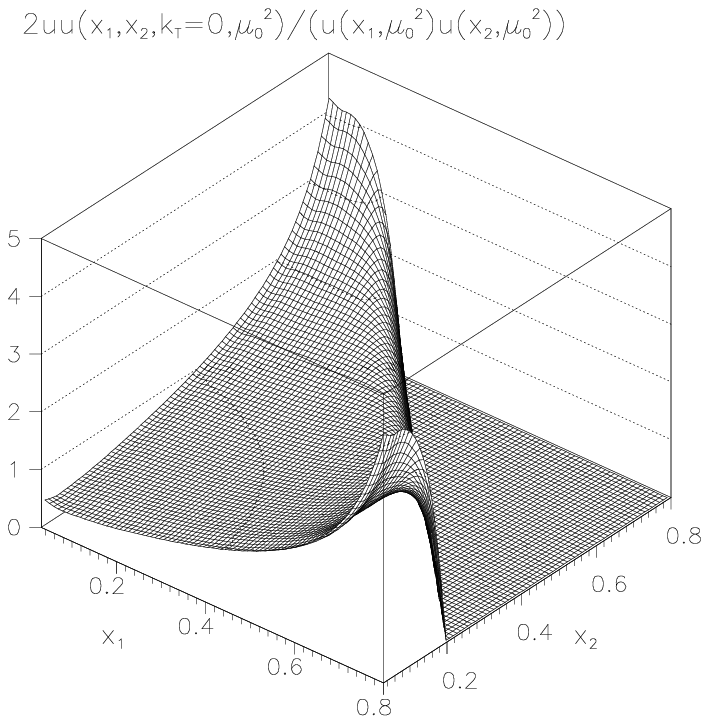}
\includegraphics{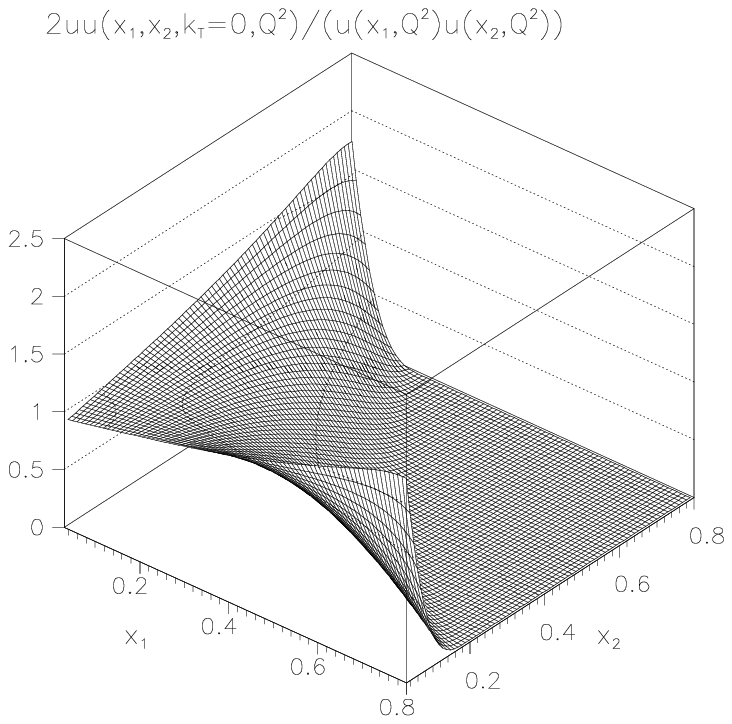}
\vskip -0.8cm
\caption{ \footnotesize Left panel: the ratio $r_2$, Eq. (\ref{ratio12u}) 
evaluated at the scale $\mu_0^2$. Right panel: same of the left panel but at 
the scale $Q^2 = 10$ GeV$^2$.}
\label{5}
\end{figure}
\noindent

\section{pQCD evolution of the calculated dPDFs}
\noindent
A fundamental point, discussed e.g. in  Ref. \cite{noij1}, is the 
analysis of the pQCD evolution of  model calculations of dPDFs.
This
procedure is essential to relate 
CQM predictions with present and future experimental analyses. 
For the moment
being,  the pQCD evolution of
dPDFs is known only for the longitudinal momentum dependence, which means
$k_\perp=0$, and using the same energy scale for both the acting
partons. In these case,  the evolution equations are obtained as a proper
generalization of the usual  Dokshitzer-Gribov-Lipatov-Altarelli-Parisi
(DGLAP)  ones (see Refs. \cite{23a,24a} for
details), defined for the evolution of PDFs. 
In the first part of this section, results will be shown for the
non-singlet sector.
In particular, the ratio $r_2$, Eq. (\ref{ratio12u}), has been calculated using
the dPDFs evaluated 
at a generic high energy scale, e.g., $Q^2 = 10$ GeV$^2$, 
using pQCD evolution.
As one can see on the right panel of 
in Fig. \ref{5}, for small values of $x_i$, e.g, close
to the LHC kinematics,  $r_2 \sim 1$. This means that  dynamical correlations, 
for valence quarks, are suppressed 
after the
evolution.
Nevertheless, in order to complete this analysis, in Ref. \cite{noij2}, new 
ratios 
sensitive to longitudinal correlations have been defined and calculated at high 
energy scale 
including
 perturbative and non 
perturbative degrees of freedom. 
To this aim let us generalize the expression of $r_2$ for different partonic 
species:

\vskip -0.4cm
\begin{align}
\label{eq:ratioab2}
 ratio_{ab} = \dfrac{F_{ab}(x_1,x_2 = 0.01, k_\perp=0; 
Q_2^2)}{a(x_1;Q_2^2)b(x_2 = 0.01;Q_2^2)}~;
\end{align}
\vskip -0.2cm
where here $a,b = {q,g,\bar q}$, $a(x;Q^2)$ and $b(x;Q^2)$  are the single 
PDFs for two given 
partons of flavor $a$ and $b$ and $Q^2_2 = 250 $ GeV$^2$.  
In such ratio, dPDFs in the numerator and PDFs in the denominator 
evolve  by means of different evolution equations.  Due to this 
feature, $ratio_{ab}$ is sensitive to non perturbative correlations, encoded in 
the 
proton wave function used to calculate dPDFs and PDFs and to the perturbative 
ones due to the difference 
in 
the evolution equations of dPDFs and PDFs. In order to disentangle these two 
different contributions to  understand which among theme could 
affect the  dPDF evaluations, the 
following ratios have been defined and calculated within the LF CQM approach:

\vskip -0.4cm
\begin{align}
 ratio_{ab}^P = \dfrac{F_{ab}(x_1,x_2 = 0.01, k_\perp=0; 
Q_2^2)|^P}{a(x_1;Q_2^2)b(x_2 = 0.01;Q_2^2)};~~~~
ratio_{ab}^{NP} = \dfrac{F_{ab}(x_1,x_2 = 0.01, k_\perp=0; 
Q_2^2)}{F_{ab}(x_1,x_2 = 0.01, k_\perp=0; 
Q_2^2)|^P}
\end{align}

where here:

\vskip -0.4cm
\begin{align}
 F_{ab}(x_1,x_2 = 0.01, k_\perp=0; 
Q_2^2)|^P = \big[a(x_1;Q_2^2)b(x_2 = 0.01;Q_2^2)  \big]^{dPDF evolution}~.
\end{align}
\vskip-0.3cm

\begin{figure}[t]
\vspace{15.0cm}
\includegraphics{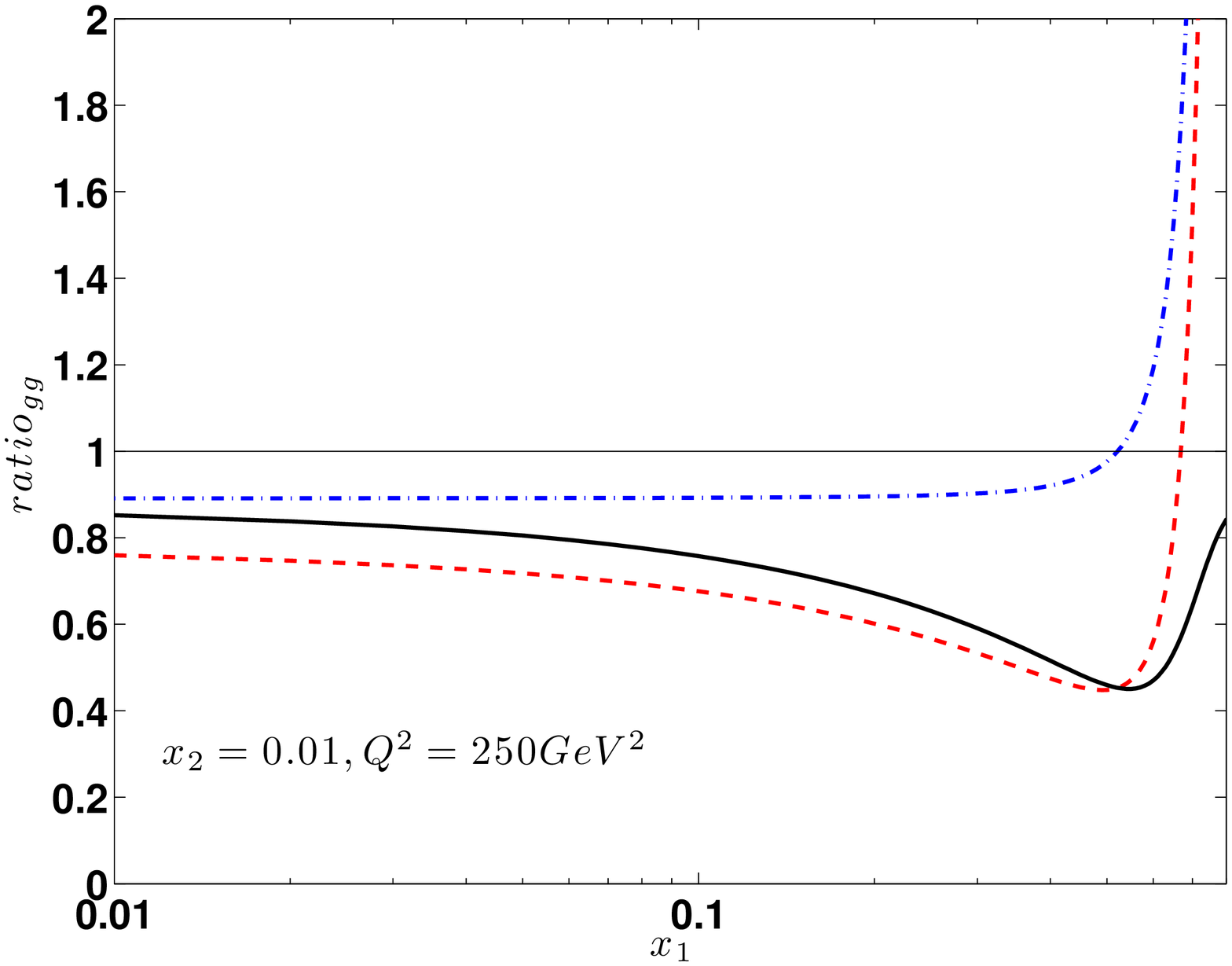}
\includegraphics{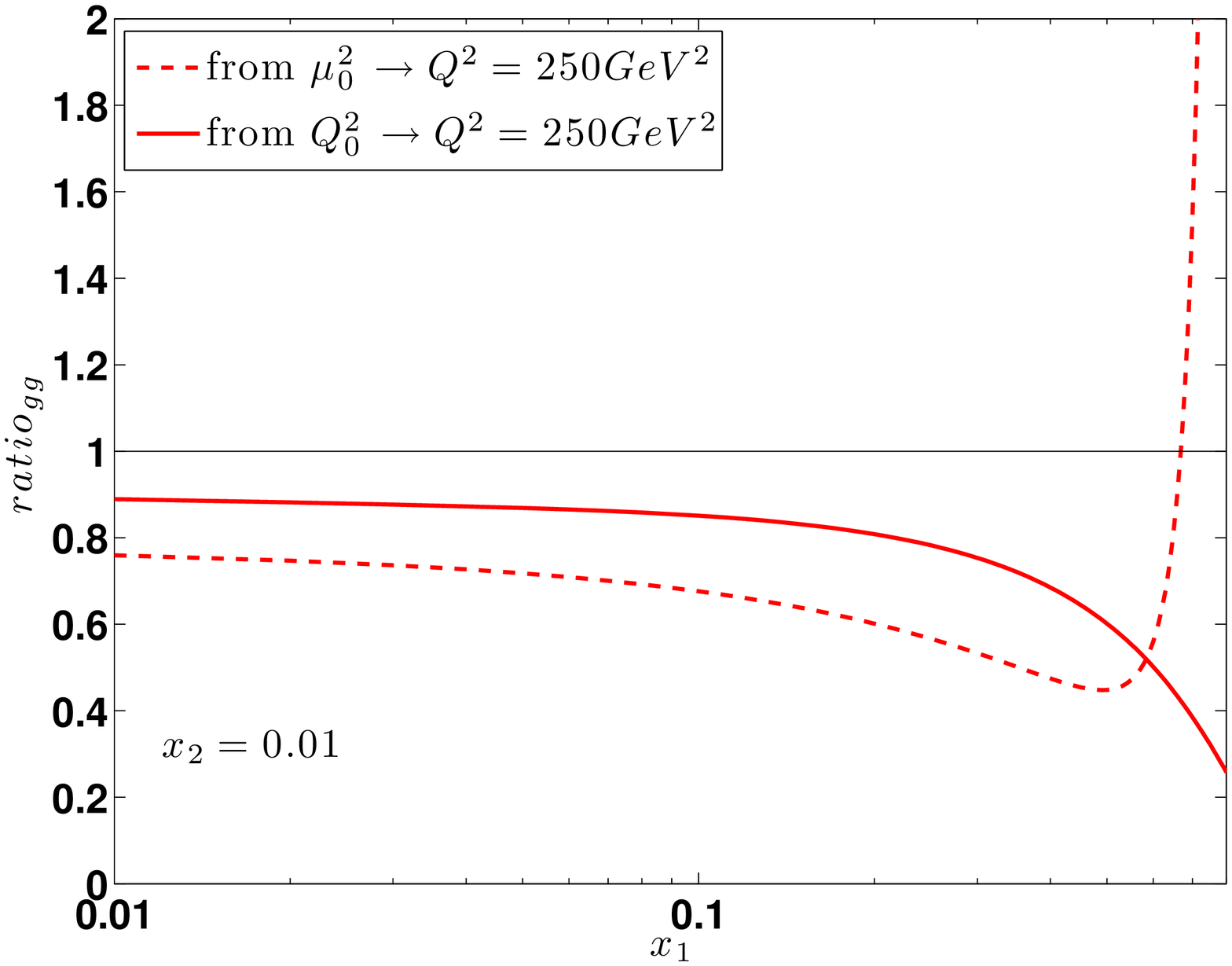}
\vskip -10cm
\caption{ \footnotesize 
Left panel: gluon gluon ratios $ratio_{gg}$ (dashed line), $ratio_{gg}^P$ 
(dot-dashed) and $ratio_{gg}^{NP}$ (continuous line) evaluated at the scale 
$Q^2_2 = 250$ GeV$^2$ and for $x_2 = 0.01$. Right panel: same of the left 
panel but including {where gluons  generated non 
perturbatively, according to  Eq. (\ref{eq:correlationQ0k}),  are taken into 
account}.} 
\label{fig:ratiogg&uVuV_mCS}
\end{figure}

The latter quantity is calculated by evolving the product of PDFs by means of 
dPDF evolution.
Due to this feature, the numerator and denominator of  $ratio_{ab}^P$  evolve 
differently in pQCD  while the input 
at the initial scale is the same, so that 
 $ratio_{ab}^P$ is sensitive to perturbative correlations. On the contrary, 
since in  ratio 
$ratio_{ab}^{NP}$ the numerator and denominator evolve within the same scheme, 
but using different input at the initial scale, it is sensitive to non 
perturbative correlations. Due to these features,
the ratio $ratio_{ab}$, shown in Fig. \ref{fig:ratiogg&uVuV_mCS} (left panel) 
for two $a=b=g$,
is particularly emblematic. The full ratio $ratio_{ab}$, Eq. 
(\ref{eq:ratioab2}), 
(dashed line), clearly influenced by both perturbative (dot-dashed line) 
and non-perturbative (continuous line) effects, is compared with those
where perturbative and 
non-perturbative correlations are disentangled,
contributing to the behavior of 
$g-g$ dPDFs at low values of $x_1$ and $x_2$.
As one can see in the left panel of Fig.  \ref{fig:ratiogg&uVuV_mCS},  for the 
gluon-gluon distribution 
 such components 
coherently interfere.
Here we report results only for gluons, being the  
highest partonic component at LHC kinematics. Furthermore,
in order to show how much these conclusions  do not 
depend on the used model, a semi factorized ansatz has been adopted in order to 
include non perturbative see quarks and gluons at a given 
initial scale. All details of the procedure are discussed in Ref. 
\cite{noij2}.
We assume that at a given initial scale $Q_0> \mu_0$ one has:

\vskip -0.5cm
\begin{align}
\label{eq:correlationQ0k}
 F_{uu}(x_1,x_2, k_\perp &=0, Q^2_0) \sim F_{u_V u_V}(x_1,x_2, k_\perp = 0, 
Q^2_0)+\left\{[u_V(x_1,Q_0^2) \bar u(x_2,Q_0^2)+   \right.
\\
& +    
 \left. \bar u(x_1,Q_0^2) u_V(x_2,Q_0^2) ] + 
\bar u(x_1,Q_0^2) \bar u(x_2,Q_0^2)\right\} (1-x_1-x_2)^n 
\theta(1-x_1-x_2)\,. \nonumber 
\end{align}
\vskip-0.3cm

where in the above expression, $F_{u_V u_V}(x_1,x_2, k_\perp = 0, 
Q^2_0)$ and $u_V(x_1,Q_0^2)$
are
 the dPDFs and PDFs respectively calculated by means of the LF CQM approach 
obtained evolving 
the same quantities from $\mu_0^2$ to $Q_0^2$ and $\bar u(x_2,Q_0^2)$ is the PDF 
evaluated 
through to the MSTW2008 parametrization \cite{MSTW}, see Ref. \cite{noij2} 
for further details. In particular, within this choice, $Q_0^2 = 1$ GeV$^2$.
As one can see in the right panel of Fig. \ref{fig:ratiogg&uVuV_mCS}, where 
$ratio_{gg}$ has been calculated starting the evolution
from  $\mu^2_0$ and $Q_0^2$,
conclusions arisen from the precedent analysis, see left panel of Fig. 
\ref{fig:ratiogg&uVuV_mCS},   are confirmed. 
For gluon-gluon  the correlations induced at low-$x$ still contain a 
specific 
sign of the correlations introduced in the valence sector and this is due to 
the 
presence of the valence component in the quark-singlet sector in the evolution 
procedure. The strength of the correlations seems to become smaller but they 
are 
still sizable and should not be neglected.

\section{Calculation of  $\sigma_{eff}$}

As previously mentioned, an important quantity, relevant for the experimental 
analyses of DPS is the
 effective cross section, $\sigma_{eff}$ whose
 expression 
in terms of PDFs and dPDFs,
has been  presented in Ref. \cite{noiPLB}. 
In order to discuss the main features of $\sigma_{eff}$, we
restrict the analysis to the zero rapidity 
region $y=0$ and show 
results,  within the LF approach, 
in the left panel of Fig. \ref{fff}. Here the latter quantity has been 
calculated at $Q^2 = 250$ GeV$^2$ for  a sea and a valence quarks. 
It is worth to notice that
the three old experimental extractions of
$\sigma_{eff}$ from data, Refs.
\cite{afs,data0,data2}, lie
in the obtained range of values of the calculated  $\sigma_{eff}$.
It is fundamental to remark that, at the hadronic scale, as demonstrated in 
Ref. \cite{noice}, if Melosh rotations would neglected, the average value of 
$\sigma_{eff}$ would change by a factor 2, making the impact of relativistic 
effects in $\sigma_{eff}$ very strong. 
In the same figure, as one can see, DPCs generate a strong $x_i$ dependence of 
$\sigma_{eff}$ in the valence region. Such feature, as already discussed,
is related to the combined 
effect of pQCD evolution correlations and non perturbative ones.
In order to provide  a fully model independent analysis of the $x$ dependence 
of $\sigma_{eff}$, in Ref. \cite{noiads}, such quantity has been calculated 
through an evaluation of dPDFs based on the AdS/QCD correspondence, see Refs. 
\cite{ads1,ads2,ads3} for fundamental details. In 
particular the authors found that, at the hadronic scale, the 
$x$ 
dependence of $\sigma_{eff}$  is comparable with the one found within the LF 
approach. Moreover, in order to emphasize such feature, in the right panel of 
Fig. 
\ref{fff}, the ratio $
{\sigma_{eff}(x_1,x_2,\mu_0^2)}/{\sigma_{eff}(x_1= 
10^{-3},x_2,\mu_0^2)   }$ has been plotted.
As one can see, such ratio  strongly depends on 
$x_2$ in the valence region, at the variance of the case in which longitudinal 
correlations were neglected.

\begin{figure}[t]
\begin{center}
\epsfig{file=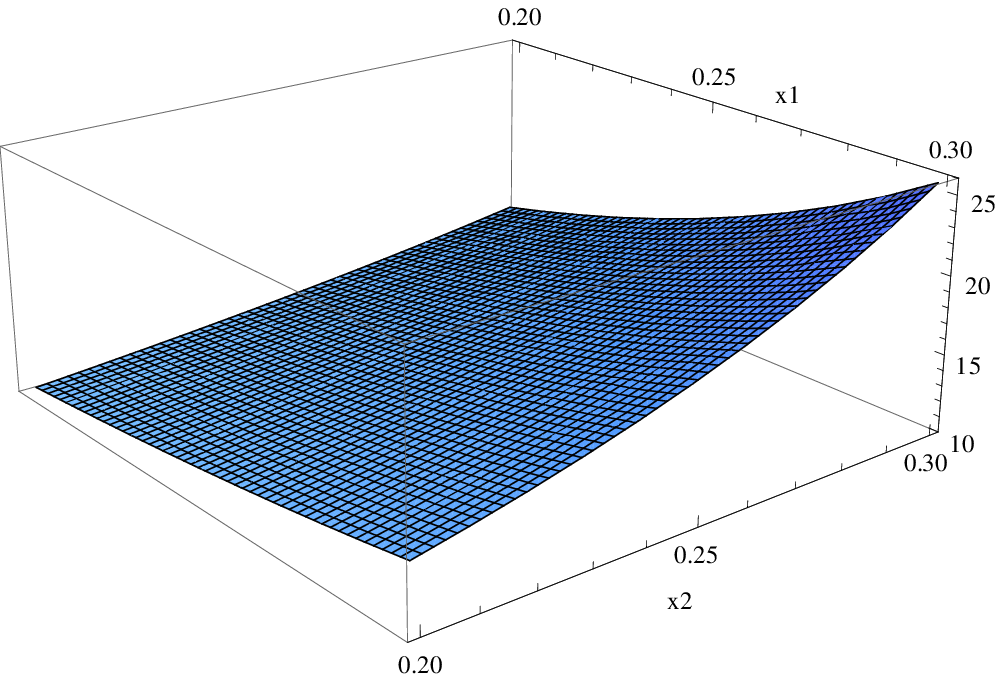,
width=6.cm
}
\epsfig{file=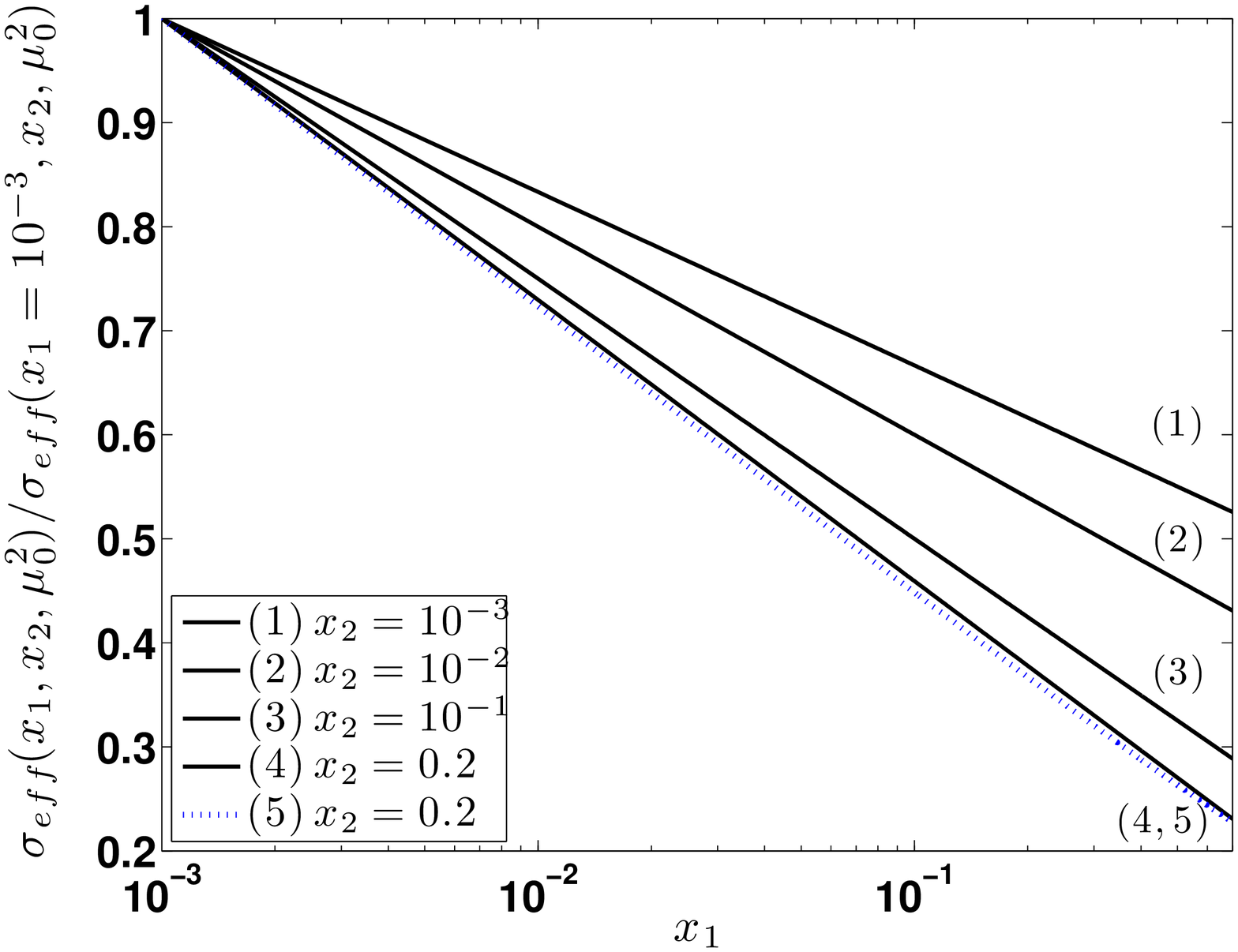, 
width=6.cm
}
\end{center}
\caption{Left panel: $ \sigma_{eff}(x_1,x_2,Q^2)$ for the values of $x_1, x_2$
measured in Ref. \cite{afs} evaluated for a valence quark and a sea quark. 
Right panel: The ratio $
{\sigma_{eff}(x_1,x_2,\mu_0^2)}/{\sigma_{eff}(x_1= 
10^{-3},x_2,\mu_0^2)   }$   as a function of as a function 
of $x_1$ at fixed $x_2 = 0.001, 0.01, 0.1, 0.2.$}
\label{fff}
\end{figure}

\section{Conclusions}

In this work, dPDFs have been calculated within  a fully Poincar\'e covariant 
constituent quark model approach, the Light-Front one, fulfilling the essential 
properties of dPDFs, see
Ref. \cite{noij1}. The main goal of 
this first analysis was to establish the role of DPCs in dPDFs. In particular, 
the factorization of dPDFs, at the hadronic scale of the model, in the 
$x_1-x_2$ and $(x_1,x_2)-k_\perp$ dependences is violated, as already found in 
previous analyses c.f. Refs. \cite{man,noi1}. In order to compare 
our 
results with actual and future experimental studies, the pQCD evolution of the 
calculated dPDFs  has been deeply investigated as discussed in 
Refs. \cite{noij1,noij2} where, at very high energy scales, like the 
experimental 
ones, it was found that longitudinal correlations may survive also at small 
values of $x_i$. Furthermore, as 
also 
explained in Ref. \cite{noij2}, where also 
non perturbative degrees of freedom have been taken into account in the 
analysis, there are different source of longitudinal correlations, the 
perturbative one, induced by the pQCD evolution scheme and the non perturbative 
ones due to the 
model used to describe the proton wave function, used to calculate the dPDFs. 
For example, in the gluon gluon case, such correlations coherently interfere in 
the small $x_i$ region at very high energy scales making theme sizable.
However, since dPDFs can not be directly 
measured in DPS,  within this model calculation approach, the so called 
$\sigma_{eff}$, a quantity experimentally studied to estimate the DPS cross 
section, has been calculated.  The latter one could be 
very important from the theoretical and experimental point of view to obtain 
information on partonic structure of the proton being 
sensitive to DPCs. In particular, signals of DPCs in $\sigma_{eff}$, have been 
found in its $x_i$ dependence in the valence region also at high 
energy scales, for details see Ref. \cite{noiPLB}. Furthermore $\sigma_{eff}$  
has been also calculated within the AdS/QCD correspondence and also in this 
case, an important  $x_i$ dependence  has been found, in 
particular in the valence region. Since, the role of correlations is crucial to 
grasp new information on the partonic structure of the proton, relativistic 
effects in the calculation of dPDFs have been studied in order to identify 
model independent correlations in the $(x_1,x_2)-k_\perp$ dependence of 
dPDFs. In 
particular, as deeply discussed in Ref. \cite{noice}, the role of Melosh 
rotations, essential feature of the LF approach, introduce non negligible 
correlations in such dependence. Furthermore new studies for the extraction of 
the proton dPDFs from $pA$ collisions are under investigations, see Ref. 
\cite{salv}.
We conclude that the investigation of dPDFs, also trough the analysis of 
$\sigma_{eff}$, could unveil new and interesting information on the 3D partonic 
structure of the proton.
I thank Sergio Scopetta, Marco Traini, Vicente Vento and Federico Alberto 
Ceccopieri for a nice and fruitful collaboration on this subject. This work was 
supported in part by the Mineco under contract FPA2013-47443-C2-1-P and 
SEV-2014-0398.

}

\end{document}